\documentclass[sigconf,nonacm]{acmart}
\usepackage{xspace}
\usepackage{xcolor}
\usepackage{pgfplots}
\usepackage{subcaption}
\usepackage[capitalise,nameinlink]{cleveref}
\pgfplotsset{compat=1.16}

\usetikzlibrary{shapes.geometric}
\usetikzlibrary{arrows}
\usetikzlibrary{decorations.pathmorphing}
\usetikzlibrary{quotes}

\tikzset{every edge quotes/.style =
          { fill = white}}
          
\makeatletter
\pgfdeclareshape{document}{
\inheritsavedanchors[from=rectangle]
\inheritanchorborder[from=rectangle]
\inheritanchor[from=rectangle]{center}
\inheritanchor[from=rectangle]{north}
\inheritanchor[from=rectangle]{south}
\inheritanchor[from=rectangle]{west}
\inheritanchor[from=rectangle]{east}
\inheritanchor[from=rectangle]{north east}
\backgroundpath{
\southwest \pgf@xa=\pgf@x \pgf@ya=\pgf@y
\northeast \pgf@xb=\pgf@x \pgf@yb=\pgf@y
\pgf@xc=\pgf@xb \advance\pgf@xc by-10pt
\pgf@yc=\pgf@yb \advance\pgf@yc by-10pt
\pgfpathmoveto{\pgfpoint{\pgf@xa}{\pgf@ya}}
\pgfpathlineto{\pgfpoint{\pgf@xa}{\pgf@yb}}
\pgfpathlineto{\pgfpoint{\pgf@xc}{\pgf@yb}}
\pgfpathlineto{\pgfpoint{\pgf@xb}{\pgf@yc}}
\pgfpathlineto{\pgfpoint{\pgf@xb}{\pgf@ya}}
\pgfpathclose
\pgfpathmoveto{\pgfpoint{\pgf@xc}{\pgf@yb}}
\pgfpathlineto{\pgfpoint{\pgf@xc}{\pgf@yc}}
\pgfpathlineto{\pgfpoint{\pgf@xb}{\pgf@yc}}
\pgfpathlineto{\pgfpoint{\pgf@xc}{\pgf@yc}}
}
}
\makeatother
          
\tikzstyle{doc}=[%
draw,
thick,
align=center,
color=black,
shape=document,
minimum width=20mm,
minimum height=28.2mm,
shape=document,
inner sep=2ex,
]

\newcommand{\gearmacro}[5]{%
\foreach \i in {1,...,#1} {%
  [rotate=(\i-1)*360/#1]  (0:#2)  arc (0:#4:#2) {[rounded corners=1.5pt]
             -- (#4+#5:#3)  arc (#4+#5:360/#1-#5:#3)} --  (360/#1:#2)
}}  

\newcommand{\chaptergear}{\begin{tikzpicture}
   \fill[gray] (0,0) circle (2cm);
   \draw[thick,rotate=12,fill=gray] \gearmacro{8}{2}{2.4}{20}{2};
   \draw[thick,fill=white] (0cm,0cm) circle(1.35cm);
 \end{tikzpicture}}
 
 \newsavebox\gearbox
\sbox\gearbox{\raisebox{-0.4em}{\resizebox{!}{1.5em}{\chaptergear}}}
  
\AtBeginDocument{%
  }

\setcopyright{acmcopyright}
\copyrightyear{2024}
\acmYear{2024}
\acmDOI{XXXXXXX.XXXXXXX}

\begin{document}

\title{Wasm SpecTec: Engineering a Formal Language Standard}

\author{Joachim Breitner}
\affiliation{%
  \institution{Independent}
  \country{Germany}}
\email{mail@joachim-breitner.de}

\author{Philippa Gardner}
\affiliation{%
  \institution{Imperial College London}
  \country{United Kingdom}}
\email{p.gardner@imperial.ac.uk}

\author{Jaehyun Lee}
\affiliation{%
  \institution{KAIST}
  \country{South Korea}}
\email{99jaehyunlee@kaist.ac.kr}

\author{Sam Lindley}
\affiliation{%
  \institution{The University of Edinburgh}
  \country{United Kingdom}}
\email{Sam.Lindley@ed.ac.uk}

\author{Matija Pretnar}
\affiliation{%
  \institution{University of Ljubljana}
  \country{Slovenia}}
\email{matija.pretnar@fmf.uni-lj.si}

\author{Xiaojia Rao}
\affiliation{%
  \institution{Imperial College London}
  \country{United Kingdom}}
\email{xiaojia.rao19@imperial.ac.uk}

\author{Andreas Rossberg}
\affiliation{%
  \institution{Independent}
  \country{Germany}}
\email{rossberg@mpi-sws.org}

\author{Sukyoung Ryu}
\affiliation{%
  \institution{KAIST}
  \country{South Korea}}
\email{sryu.cs@kaist.ac.kr}

\author{Wonho Shin}
\affiliation{%
  \institution{KAIST}
  \country{South Korea}}
\email{new170527@kaist.ac.kr}

\author{Conrad Watt}
\affiliation{%
  \institution{University of Cambridge}
  \country{United Kingdom}}
\email{conrad.watt@cl.cam.ac.uk}

\author{Dongjun Youn}
\affiliation{%
  \institution{KAIST}
  \country{South Korea}}
\email{f52985@kaist.ac.kr}

\newcommand{\dslname}{SpecTec\xspace}
\newcommand{\todo}[1]{{\color{red}#1}}

\crefformat{section}{#2\S{}#1#3}
\crefname{figure}{Figure}{Figures}
\crefname{lstlisting}{Listing}{Listings}
\crefname{algorithm}{Algorithm}{Algorithms}
\crefname{table}{Table}{Tables}
\Crefname{line}{Line}{Lines}
\crefname{line}{line}{lines}

\begin{abstract}
WebAssembly (Wasm) is a low-level bytecode language and virtual machine, intended as a compilation target for a wide range of programming languages,  which is seeing increasing adoption across
diverse ecosystems.  As a young technology, Wasm continues to evolve --- it reached version 2.0 last year and another major update is expected soon.

For a new feature to be standardised in Wasm, four key artefacts
must be presented: a formal (mathematical) specification of the feature,  an accompanying prose pseudocode description, an implementation in the official reference
interpreter, and a suite of unit tests.
This rigorous process helps to avoid errors in the design and implementation of new Wasm features, and Wasm's distinctive formal specification in particular has facilitated machine-checked proofs of various correctness properties for the language.
 However, manually crafting all of these artefacts requires expert knowledge combined with
repetitive and tedious labor, which is a burden on the language's standardization process and authoring of the specification.

This paper presents \emph{Wasm \dslname},
a technology to express the \emph{formal} specification of Wasm through a \emph{domain-specific language}. This DSL allows all of Wasm's currently handwritten specification artefacts to be error-checked and generated automatically from a single source of truth, and is designed to be easy to write, read, compare, and review.
We believe that \emph{Wasm \dslname}'s automation and meta-level error checking will significantly ease the current burden of the language's specification authors.
We demonstrate the current capabilities of Wasm \dslname by showcasing its
proficiency in generating various artefacts, and describe our work towards replacing the manually written
official Wasm specification document with specifications generated by Wasm \dslname.
\end{abstract}

\maketitle

\begin{figure*}[t]
\centering
\begin{tikzpicture}

\draw (9,2) node[draw,doc, minimum height=.8cm,minimum width=3cm]
(FILES) {SpecTec Files (DSL)};

\draw (5,1) node[draw,rectangle, rounded corners, minimum height=.8cm,minimum width=3cm] (EXT) {\begin{tabular}{c}External\\Representation\end{tabular} (EL)};

\draw (5,-1) node[draw,rectangle, rounded corners, minimum height=.8cm,minimum width=3cm] (INT) {\begin{tabular}{c}Internal\\Representation\end{tabular} (IL)};

\draw (5,-3) node[draw,rectangle, rounded corners, minimum height=.8cm,minimum width=3cm] (ALG) {\begin{tabular}{c}Algorithmic\\Representation\end{tabular} (AL)};

\draw (-3,-4.8) node[draw,doc, minimum height=.8cm,minimum width=3cm] (FORM) {Formalism (LaTeX)};

\draw (1,-4.8) node[draw,doc, minimum height=.8cm,minimum width=3cm] (PROSE) {Prose (Sphinx)};

\draw (5,-4.8) node[draw,rectangle, minimum height=.8cm,minimum width=3cm] (ARINT) {AL Interpreter};

\draw (11,-4.5) node[draw,doc, minimum height=.8cm,minimum width=3cm] (MECH) {\begin{tabular}{c}Mechanisations\\(Coq/Isabelle/Agda/Lean)\end{tabular}};

\draw (5,-6.4) node[draw,doc, minimum height=.8cm,minimum width=3cm] (TEST) {Test Suite / Fuzzer};

\draw (11,-6.4) node[draw,doc, minimum height=.8cm,minimum width=3cm] (PROOF) {Proof Obligations};

\draw [line width=0.4mm, -stealth] (FILES.south) to[out=270,in=0, distance=0.5cm] node[below right, yshift=1.0ex]{\hspace{0.0em} \usebox{\gearbox} parsing} (EXT.east);
\draw [line width=0.6mm, -stealth] (EXT.south) to node[right, yshift=0.7ex]{\hspace{0.0em} \usebox{\gearbox} checking, elaboration} (INT.north);
\draw [line width=0.6mm, -stealth] (INT.south) to node[right, yshift=0.7ex]{\hspace{0.0em} \usebox{\gearbox} animation} (ALG.north);
\draw [line width=0.6mm, -stealth] (ALG.south) -- (ARINT.north);

\draw [line width=0.6mm, -stealth, dashed] (TEST.north) -- (ARINT.south);
\draw [line width=0.6mm, -stealth,dashed] (PROOF.north) -- (MECH.south);

\draw [line width=0.6mm, -stealth] (ALG.west) to[out=180,in=90, distance=1.5cm,"\usebox{\gearbox} prose backend", pos=.8] (PROSE.north);

\draw [line width=0.6mm, -stealth] (INT.east) to[out=0,in=90, distance=3cm,"\usebox{\gearbox} prover backends", pos=.7] (MECH.north);

\draw [line width=0.6mm, -stealth] (EXT.west) to[out=180,in=90, distance=4cm, "\usebox{\gearbox} LaTeX backend"] (FORM.north);

\draw [line width=0.6mm, -stealth] (INT.north west) to[out=135,in=225, distance=1cm] node[left]{\usebox{\gearbox} transforms} (INT.south west);

\draw [line width=0.6mm, -stealth] (INT.east) to[out=0,in=45, distance=2.5cm] (TEST.north east);

\end{tikzpicture}
\caption{An overview of Wasm \dslname}
\label{fig:structure}
\vspace{-0.0\baselineskip}
\end{figure*}
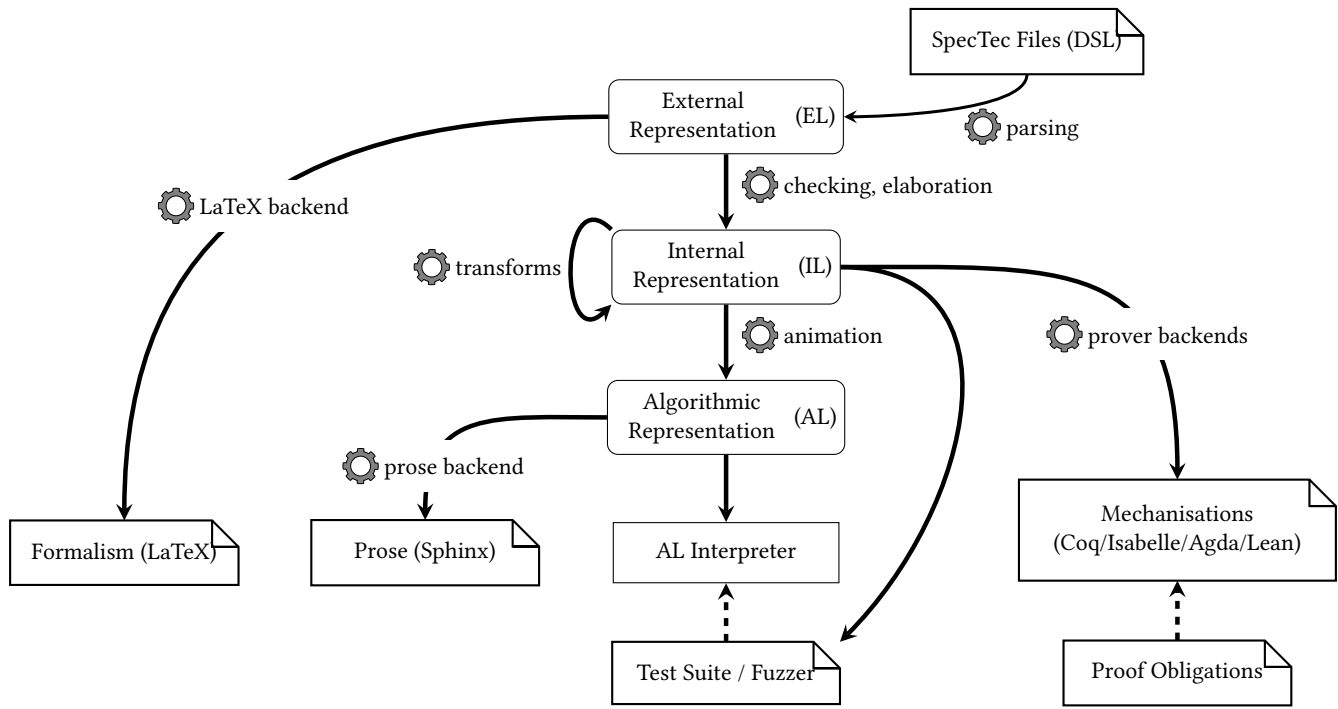

\section{Introduction}
\label{sec:intro}

WebAssembly (Wasm) is a low-level bytecode language and virtual machine~\cite{wasm-pldi17}.  Initially introduced to
allow efficient compilation and execution of a larger variety of programming
languages on the Web platform, it has since been adopted across a broad range
of ecosystems, such as cloud and edge computing~\cite{lucet, cloudflare},
mobile and embedded systems~\cite{wasm-embedded}, IoT~\cite{wasm-iot}, and
blockchains~\cite{wasm-blockchain}.

Following its initial release, there has been growing demand for the
integration of new language features into Wasm. The version of Wasm initially
supported by browsers in 2017, referred to by its designers as a ``Minimum
Viable Product'' (MVP)~\cite{wasm-mvp}, served as a simple compilation target
for languages like C/C++ and Rust. However, the performance of
programs compiled with the MVP is suboptimal, and a few key features such as shared
memory concurrency and exception handling are not supported by the MVP. Other
languages involving runtime-managed memory such as Java
cannot be compiled without significant performance
or usability compromises, for example, because the Wasm MVP cannot easily
express the stack walking strategies used by the garbage collectors of
these languages' runtimes. Several proposals, including SIMD
Vector Instrutions~\cite{wasm-simd}, Exception Handling~\cite{wasm-exc}, Garbage Collected Types~\cite{wasm-gc}, and Threads~\cite{wasm-threads}
are being developed in Wasm to alleviate these problems.

For a feature to be standardised in Wasm, four key artefacts must be presented
to the W3C Wasm Community Group~\cite{wasm-w3c}:
\begin{itemize}
\item a \emph{formal specification} for the feature in the form of mathematical rules, written in LaTeX;
\item a \emph{prose pseudocode} description of the feature's behaviour, written in reStructuredText markup;
\item an implementation of the feature in the Wasm \emph{reference interpreter}, written in OCaml; and
\item a suite of \emph{unit tests} for the feature, written in (an enriched version of) the Wasm text format.
\end{itemize}
First, a formal specification significantly reduces the risk of under-defined edge cases. Indeed, if it is also accompanied by formal proofs of
appropriate correctness properties, as has been the case for the Wasm MVP,
certain risks such as type safety violations can be entirely precluded~\cite{Watt2018MechanisingAV, Watt2021Two}.
Second, the prose pseudocode description is designed to be more accessible to
non-experts, similar to other normative language specifications such as JavaScript's~\cite{ecmascript}.
Third, a reference interpreter can often be useful in situations where an optimised
implementation in a production engine is not yet available or is substantially
more complex.  Finally, the insistence on a comprehensive test suite for each
newly added feature further guarantees that various implementations of Wasm
exhibit consistent behavior.  This collective effort to include both
formal and prose specifications,
reference interpreters, and thorough test suites underscores the commitment to
precision, reliability, and compatibility within the Wasm standardization
process, and serve to reduce the risk of implementation divergence.

This meticulous process exists as a reaction to past experiences; historically Web languages have been particularly vulnerable to
issues of implementation discrepancy.  Leaving aside deliberate breaks from a Web standard by a browser vendor,
discrepancies may occur inadvertently simply due to the number of different browser implementations in existence, each one in itself consisting of several tiers of interpretation and just-in-time compilation.  Additionally, developers deploying code on the Web platform
have particuarly limited control over a Web site visitor's execution environment, amplifying the impact of
any discrepancies.  Portability is only feasible if implementations are meticulous in aligning their behaviours.

One significant challenge in Wasm's current standardisation
process~\cite{wasm-process} lies in the poor developer experience,
where the developer in question is an author of the specification artefacts described above.
The development of these
artefacts has on occasion lagged significantly behind other aspects of the standardisation process,
delaying the integration of a new proposal into the standard.
For example the highly anticipated Threads~\cite{wasm-threads} proposal has not yet been standardized in large part due to specification authoring delays.

The current Wasm specification~\cite{wasm-spec} is authored in reStructuredText,
a (somewhat cumbersome) markup language, from which both HTML and PDF documents
are generated by the Sphinx document processor \cite{sphinx}.
The formal pieces of the specification consist of embedded mathematics that must be
expressed in a (severely restricted) subset of LaTeX;
for HTML, it is rendered by MathJax \cite{mathjax}.
Wasm specification authors complain of the following significant difficulties in preparing and maintaining the specification text:\vspace{-0\baselineskip}
\begin{itemize}
\item lack of visual clarity when reading the raw source (complicating \emph{code reviews} and necessitating repeated lengthy builds);
\item absence of useful abstraction capabilities in Sphinx markup and in the available LaTeX subset
(due to the limitations of Sphinx and MathJax);
\item difficulty in interpreting LaTeX errors (because Sphinx generates one monolithic file before passing it to LaTeX, destroying line number information); and
\item no protection against misuse of definitions (e.g. wrong arguments, incorrect placement, incorrect symbol bindings).
\end{itemize}
\vspace{-0\baselineskip}

To address these challenges and improve the productivity of Wasm's standards
developers, we propose a unified domain-specific specification language and
corresponding toolchain, \emph{Wasm \dslname}.  \dslname will alleviate the
burden on developers by conducting meta-level error checking and automatically
generating the required specification artefacts.  Unlike existing
general-purpose specification languages such as Ott~\cite{ott},
PLTRedex~\cite{pltredex}, Skeleton~\cite{skeleton}, the K framework~\cite{k},
or Spoofax~\cite{spoofax}, our solution is unashamedly specialised to Wasm,  both to provide a development experience tailored to the expectations and needs of Wasm's standards
community, and to pursue more ambitious analyses and generated outputs which are only tractable with this more targetted scope. 
We ultimately aim for the Wasm standards community to specify all current and future Wasm features using \dslname and replace the manually authored
artefacts necessary for Wasm's standardization process with our generated artefacts, enhancing the  standardization process' efficiency and reliability.
Our in-development \dslname toolchain is available publicly~\cite{spectec}.

\section{Wasm \dslname}
\label{sec:structure}

An overview of Wasm \dslname is illustrated in Figure~\ref{fig:structure}.
The Wasm specification is primarily concerned with defining the binary format, type system, and runtime behaviour of Wasm.
With \dslname, an author will write these definitions in our \emph{Domain-Specific Language (DSL)}, which the Wasm \dslname toolchain accepts as input.
This input is parsed as the \textit{External Language (EL)} representation and processed into further representations,
namely the \textit{Internal Language (IL)} representation,
and the \textit{Algorithmic Language (AL)} representation.
Our various backends use these representations to produce the previously-described output artefacts, as well as \textit{mechanised} definitions in \textit{interactive theorem provers} suitable for machine-checked proofs about the language semantics~\cite{Watt2018MechanisingAV, Watt2021Two}.

\begin{figure}[t]
  \centering
  \begin{subfigure}{\columnwidth}
    \centering
    \includegraphics[width=\columnwidth]{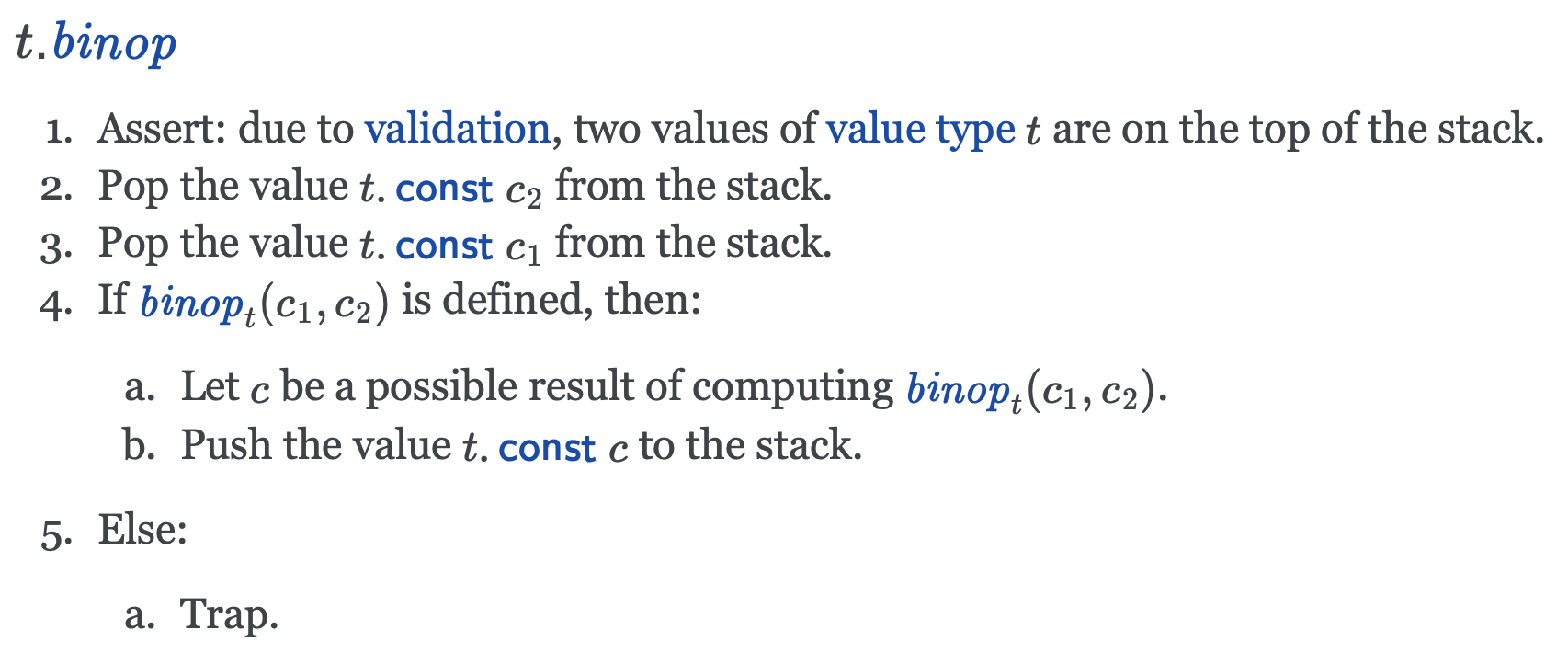}
    \subcaption{Prose specification}
\vspace*{1em}
  \end{subfigure}

  \begin{subfigure}{\columnwidth}
    \centering
    \includegraphics[width=\columnwidth]{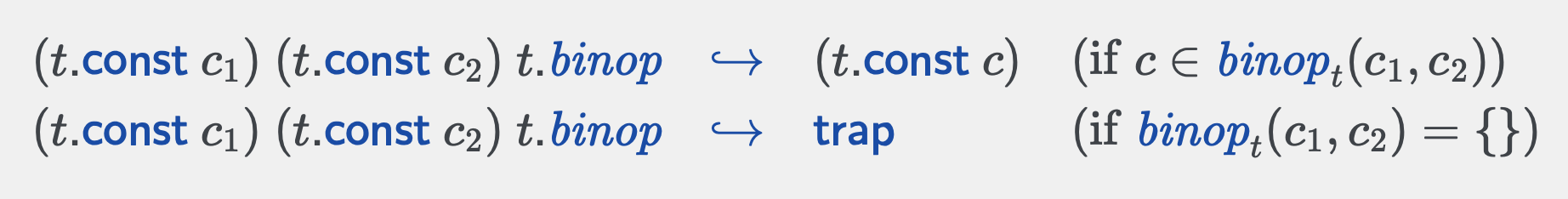}
    \subcaption{Formal specification}
  \end{subfigure}
\caption{The binary operator semantics in the specification}
\label{fig:spec}
\end{figure}

As discussed in \cref{sec:intro}, the Wasm specification is currently written in a mixture of
reStructuredText and raw LaTeX.  Figure~\ref{fig:spec} presents the execution
semantics of the $t.\mbox{\emph{binop}}$ binary operator instruction in the Wasm specification~\cite[Section 4.4]{wasm-spec}, as rendered today.
Figure~\ref{fig:spec}(a) shows
the prose pseudocode describing the five execution steps,
and Figure~\ref{fig:spec}(b) shows the mathematics in the gray box for the corresponding
operational semantics.
\dslname aims to provide a significantly better developer experience without compromising on the fidelity of the rendered specification.
Moving forward, we will provide a
comprehensive breakdown of each step.

\dslname's DSL and EL are intended to closely mirror an ASCII representation of the syntactic constructs used in Wasm's formal specification.
Figure~\ref{fig:dsl} gives a DSL definition of the runtime semantics of Wasm's binary arithmetic operator,
where {\small\verb!$binop! }is a separately defined auxiliary function.
Crucially, all definitions and variables in the parsed EL are ``type-checked'' so that ill-formed definitions can be detected.
For example, if a specification author misses the final argument of 
{\small\verb!$binop!} as in {\small\verb!$binop(binop, nt, c_1)!},
then an arity-mismatch error will be raised.
From the EL, \dslname can directly produce the LaTeX-based formal specification, which is intended to replace the current specification's handwritten definitions.
Figure~\ref{fig:latex} is an excerpt from the PDF generated from the LaTeX
translated from Figure~\ref{fig:dsl}.  Compare this to the original handwritten LaTeX in Figure~\ref{fig:spec}.

To produce the other artefacts mentioned in Section~\ref{sec:intro},
the \dslname definitions are processed further.
First, the EL is elaborated into the IL, suitable for deep analysis and transformation.
Among other things, types and multiplicities of variables are inferred and annotated in the IL, mutually recursive definitions are identified, and implicit upcasts are made explicit and disambiguated.
As Figure~\ref{fig:structure} illustrates, the IL can undergo internal transformations to meet the needs of subsequent backends.
For example, in the EL expressions are modelled as
relations that can fail or can denote multiple values,  whereas various theorem prover
backends require that expressions must be purely functional, i.e.\ must denote exactly
one value given values for all free variables.
Figure~\ref{fig:lean} is an excerpt from the code generated for the Lean theorem prover~\cite{Moura2015TheLT}.
We are currently working on similarly generating code for Coq, Isabelle, and Agda.

The operational semantics of Wasm described in the IL is further transformed into the more restricted AL,
which does not allow arbitrary relational definitions and
enforces an algorithmic order of evaluation.
The problem of transforming a relational definition into an
executable, algorithmic one is known as \textit{animation}~\cite{animate}.
At its core, the process of animation involves performing a dataflow analysis on a relational definition to infer which equations of the relational definition should be interpreted as binding new variables, and ensures that these binding definitions can be ordered such that each binding definition only depends on prior animated definitions.

\begin{figure}[t]
\small
\begin{verbatim}
rule Step_pure/binop-val:
  (CONST nt c_1) (CONST nt c_2) (BINOP nt binop) ~> (CONST nt c)
  -- if $binop(binop, nt, c_1, c_2) = c

rule Step_pure/binop-trap:
  (CONST nt c_1) (CONST nt c_2) (BINOP nt binop) ~> TRAP
  -- if $binop(binop, nt, c_1, c_2) = epsilon
\end{verbatim}
\caption{The binary operator semantics in \dslname's DSL}
\label{fig:dsl}
\end{figure}

\begin{figure}[t]
\small
$$
\begin{array}{@{}l@{}lcl}
{[\textsc{\scriptsize E{-}binop{-}val}]} \ \
 & (\mathit{nt}.\mathsf{const}~\mathit{c}_{1})~(\mathit{nt}.\mathsf{const}~\mathit{c}_{2})~(\mathit{nt} . \mathit{binop})
 &\hookrightarrow& (\mathit{nt}.\mathsf{const}~\mathit{c}) \\
& \mbox{if}~{{{\mathit{binop}}{}}_{\mathit{nt}}}{(\mathit{c}_{1},\, \mathit{c}_{2})} = \mathit{c} \\
{[\textsc{\scriptsize E{-}binop{-}trap}]} \ \
 & (\mathit{nt}.\mathsf{const}~\mathit{c}_{1})~(\mathit{nt}.\mathsf{const}~\mathit{c}_{2})~(\mathit{nt} . \mathit{binop})
 &\hookrightarrow& \mathsf{trap} \\
& \mbox{if}~{{{\mathit{binop}}{}}_{\mathit{nt}}}{(\mathit{c}_{1},\, \mathit{c}_{2})} = \epsilon
\end{array}
$$
\vspace*{-1em}
\caption{The binary operator semantics in a generated PDF}
\label{fig:latex}
\end{figure}

\begin{figure}[t]
\footnotesize
\begin{verbatim}
  | binop_val  (binop : Binop_numtype) (c : C_numtype) (c_1 : C_numtype)
               (c_2 : C_numtype) (nt : Numtype) : 
    ((«$binop» (binop, nt, c_1, c_2)) == [c]) -> 
    (Step_pure ([(Admininstr.CONST (nt, c_1)),
                 (Admininstr.CONST (nt, c_2)),
                 (Admininstr.BINOP (nt, binop))],
                [(Admininstr.CONST (nt, c))]))

  | binop_trap (binop : Binop_numtype) (c_1 : C_numtype)
               (c_2 : C_numtype) (nt : Numtype) : 
    ((«$binop» (binop, nt, c_1, c_2)) == []) -> 
    (Step_pure ([(Admininstr.CONST (nt, c_1)),
                 (Admininstr.CONST (nt, c_2)),
                 (Admininstr.BINOP (nt, binop))],
                [Admininstr.TRAP]))
\end{verbatim}
\caption{The binary operator semantics in generated Lean}
\label{fig:lean}
\end{figure}

{\sloppypar
Figure~\ref{fig:al} shows how the declarative specification of the
binary operator semantics from Figure~\ref{fig:dsl} is translated to the algorithmic version.
Note that the implicit conditions guaranteed by the Wasm validation phase
are explicitly added as assertions such as
{\small\verb!AssertI(TopValueC(NameE(nt)))!}.
Interestingly, two equality expressions are translated to different AL constructs.
Because the equality ``{\small\verb!if $binop(binop, nt, c_1, c_2) = c!}'' is a binding
that binds the result of the {\small\verb!$binop!} call to {\small\verb!c!}, it is translated as
a {\small\verb!let!} instruction:
}

\smallskip
{\small
\begin{verbatim}
    LetI(ListE([NameE(c)]),
         AppE(binop, [NameE(binop), NameE(nt),
                      NameE(c_1), NameE(c_2)]))
\end{verbatim}
}
\smallskip

\noindent
On the contrary, since ``{\small\verb!if $binop(binop, nt, c_1, c_2) = epsilon!}''
is an equality check which is inferred to bind no new variable, it is translated as a conditional expression:

\smallskip
{\small
\begin{verbatim}
    CompareC(is, AppE(binop, [NameE(binop), NameE(nt),
                              NameE(c_1), NameE(c_2)]),
             ListE([]))
\end{verbatim}
}
\smallskip

\begin{figure}[t]
\footnotesize
\begin{verbatim}
execution_of_BINOP_ainstr NameE(nt) NameE(binop):
  AssertI(TopValueC(NameE(nt)))
  PopI(ConstructE(CONST_ainstr, [NameE(nt), NameE(c_2)]))
  AssertI(TopValueC(NameE(nt)))
  PopI(ConstructE(CONST_ainstr, [NameE(nt), NameE(c_1)]))
  IfI(
    CompareC(is, LengthE(AppE(binop, [NameE(binop), NameE(nt),
                                      NameE(c_1), NameE(c_2)])), 1),
    [LetI(ListE([NameE(c)]),
          AppE(binop, [NameE(binop), NameE(nt),
                       NameE(c_1), NameE(c_2)]))
     PushI(ConstructE(CONST_ainstr, [NameE(nt), NameE(c)]))],
    [])
  IfI(
    CompareC(is, AppE(binop, [NameE(binop), NameE(nt),
                              NameE(c_1), NameE(c_2)]), ListE([])),
    [TrapI],
    [])
\end{verbatim}
\caption{The binary operator semantics in \dslname's AL}
\label{fig:al}
\end{figure}

\begin{figure}[t]
\includegraphics[width=.48\textwidth]{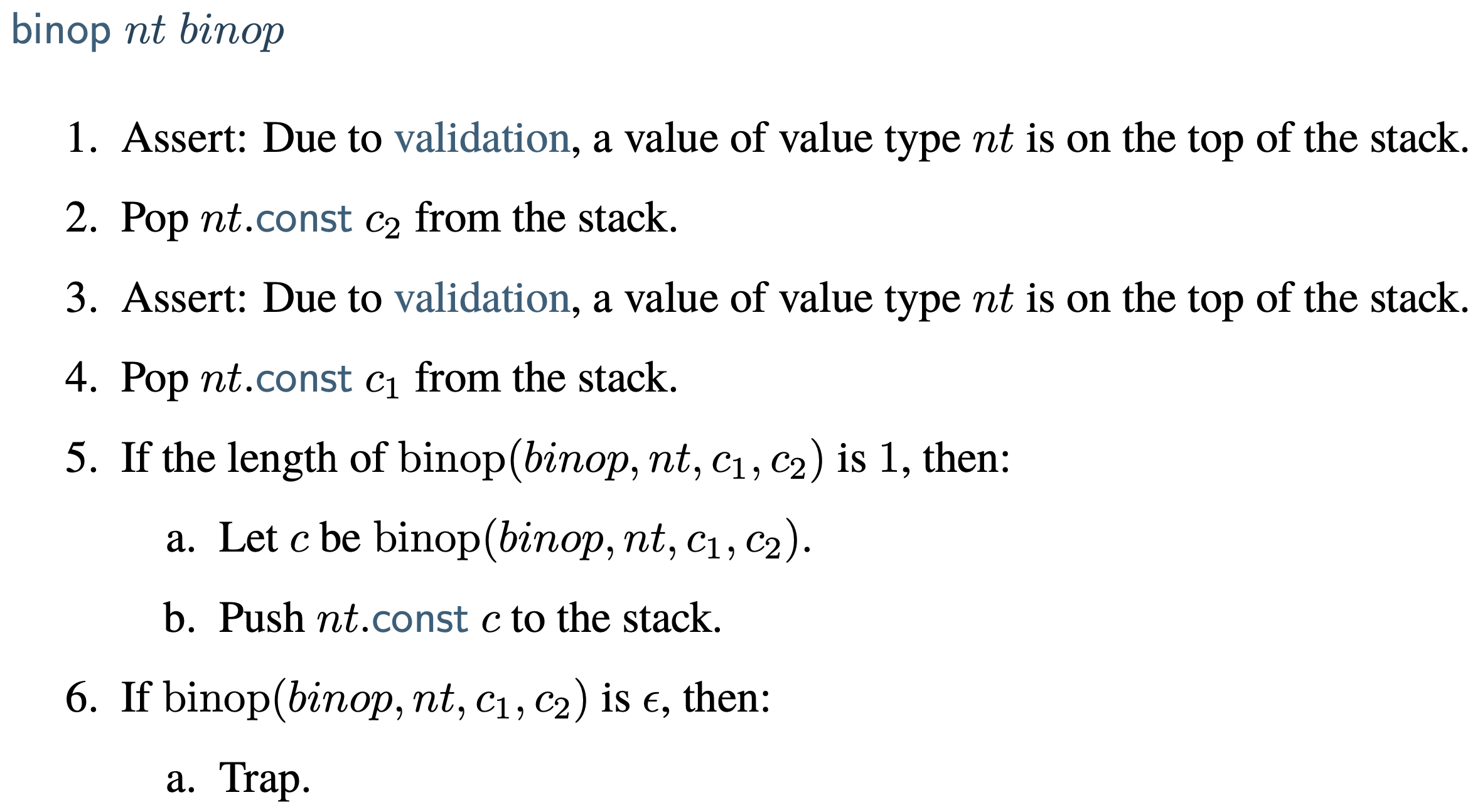}
\caption{The binary operator semantics in generated prose}
\label{fig:gen-spec}
\end{figure}

From the AL, we directly generate a prose pseudocode specification.
Figure~\ref{fig:gen-spec} shows the prose pseudocode generated from the specification in Figure~\ref{fig:al},
which is strikingly close to the original handwritten prose description in Figure~\ref{fig:spec}.
We also implement an interpreter for the AL.
By interpreting AL that represents the Wasm semantics, we indirectly obtain an interpreter for Wasm.
This technique was previously used by JISET~\cite{jiset,esmeta}
to extract an executable semantics from the ECMAScript prose that represents the official
JavaScript specification~\cite{ecmascript}, and we use it here to extract an executable semantics for Wasm, from the \dslname AL.

Currently, Wasm \dslname covers all of Wasm 2.0 except for the recently-standardized SIMD vector instructions.
Within one second, our toolchain can automatically generate both prose pseudocode and operational semantics in LaTeX
with hyperlinks and cross-references in the generated PDF document, and a Lean mechanization.
We tested the extracted Wasm semantics against the official Wasm unit test suite
on an Ubuntu machine with
a 4.0GHz Intel(R) Core(TM) i7-6700k and 32GB of RAM (Samsung DDR4 2133MHz 8GB*4).
On this machine, the extracted semantics passed all 23,778 tests (SIMD excluded) in the
test suite in 21.349 seconds.

\section{Future Plans}
\label{sec:future}

The key measure of \dslname's success is its level of adoption and ongoing support by the industrial stakeholders of the WebAssembly Community Group.
Our ultimate aim is for the normative definition of Wasm to be written using \dslname, and for all of the specification artefacts which are currently manually generated and maintained separately to be instead automatically generated from this \dslname definition as a single source of truth.

We now plan to gather feedback from industrial stakeholders, and evaluate how easily we can extend our definitions written in \dslname to cover additional features.
Wasm 3.0, an upcoming edition of the specification, may include Exception Handling~\cite{wasm-exc}, Garbage Collected Types~\cite{wasm-gc}, and Threads~\cite{wasm-threads}.
Because these features contain far more ambitious extensions to the Wasm virtual machine,
and are expected to lay the groundwork on which many future proposals will be built,
investigating the extent to which our \dslname definitions can be extended to cover these features would be the strongest possible validation of our approach.
The associated changes to the Wasm virtual machine will be wide ranging enough that we expect modifications to \dslname itself may be necessary to make it sufficiently expressive, although we are already making a best effort to anticipate the future effects of these proposals in our current design.
Succeeding in expressing all of these features as \dslname definitions would be a strong signal to industry stakeholders that \dslname can be seriously considered for official adoption.


\section{Conclusion}
\label{sec:conclusion}

We have presented Wasm \dslname, a technology for automatically
generating various artefacts from a single source of truth, written in the Wasm specification DSL.
We hope to replace the artefacts of the Wasm specification process
which are today onerously crafted by hand with those generated by Wasm \dslname.
This approach will facilitate the standardisation of future Wasm features
while improving the consistency and trustworthiness of Wasm's various specification artefacts. 
We also believe that \dslname will facilitate the production of trustworthy mechanizations
in diverse theorem provers, including Coq, Isabelle, Agda, and Lean.
While \dslname is still currently in development,
it is already capable of generating a formal specification and prose pseudocode covering all of Wasm 2.0 minus the SIMD instructions,
and an extracted Wasm semantics which passes all 23,778 applicable tests in the official Wasm test suite.
We continue to work on further backends for automatically generating unit tests and full theorem prover definitions.
We intend to use upcoming Wasm features such as Exception Handling,
Garbage-Collected Types, and Threads to further evaluate the utility and applicability of \dslname.
We acknowledge that a key challenge moving forward will be transitioning \dslname from a research tool to a robust and maintainable part of the Wasm industry standards ecosystem, and we intend to work with key Wasm industry stakeholders to achieve this goal.

\bibliographystyle{ACM-Reference-Format}
\balance
\bibliography{ref}

\end{document}